\documentclass[lettersize,journal]{IEEEtran}
\usepackage{amsmath,amsfonts}
\usepackage{algorithmic}
\usepackage{algorithm}
\usepackage{array}
\usepackage[caption=false,font=normalsize,labelfont=sf,textfont=sf]{subfig}
\usepackage{textcomp}
\usepackage{stfloats}
\usepackage{url}
\usepackage{verbatim}
\usepackage{graphicx}
\usepackage{cite}
\usepackage{caption}
\usepackage{flushend}
\usepackage{xcolor}
\usepackage{threeparttable}
\usepackage{booktabs} 
\usepackage{xurl}
\usepackage[hidelinks,breaklinks=true]{hyperref}

\begin{document}

\title{Performance Study of Serverless Workloads\\ in Confidential Virtual Machines}


\author{Rikesh Niroula, Jianchen Shan, Xiaoning Ding
\thanks{Rikesh Niroula and Xiaoning Ding are with New Jersey Institute of Technology, New Jersey, USA. Email: \{rn382,xiaoning.ding\}@njit.edu.}
\thanks{Jianchen Shan is with Hofstra University, New York, USA. Email: jianchen.shan@hofstra.edu.}}

\maketitle

\begin{abstract}
Confidential serverless computing is rapidly emerging as a critical paradigm for application domains requiring strong confidentiality guarantees, such as healthcare, finance, and machine learning. To enable this paradigm in untrusted cloud environments, Confidential Virtual Machines (CVMs) provide isolation by encrypting the entire guest memory, and thus securing serverless workloads against host-level access and interference. However, the implications of CVMs for serverless systems remain insufficiently understood. 

This paper presents an empirical study of serverless workloads in CVMs, systematically covering 
memory efficiency and runtime overhead on both warm-starts and cold-starts. Our results show that 
CVMs incur substantial memory overhead because encrypted
memory disables cross-VM page deduplication and reduces memory reclaimability, thereby limiting warm-container 
capacity under a fixed memory budget. Runtime overhead in warm-starts is workload dependent. Workloads with frequent VMEXITs, particularly idle transitions, suffer substantial slowdowns. 
These slowdowns are further amplified by common serverless deployment practice that couples vCPU allocation to 
memory size, as higher-memory configurations expose more vCPUs than some functions can use effectively.
For cold starts, the study focuses on container creation, a common and major contributor to startup latency. 
Motivated by the memory-efficiency results, we consider the deployments in which multiple 
containers of the same function are consolidated within the same CVM to improve efficiency, and 
show that selectively relaxing certain isolation mechanisms in this setting can substantially 
reduce startup overhead.
These results clarify the main performance tradeoffs of confidential serverless computing and suggest practical ways to improve efficiency and latency.

\end{abstract}

\vspace{-0.12 in}

\section{INTRODUCTION}

\IEEEPARstart{S}{erverless} computing, also known as Function-as-a-Service (FaaS), has become a popular cloud paradigm, adopted by platforms such as AWS Lambda \cite{aws_lambda}, Microsoft Azure Functions \cite{azure_functions}, and Google Cloud Functions \cite{google_cloud_functions}, due to its scalability, fine-grained billing, and reduced operational overhead.
It is well-suited for a broad range of applications, such as image processing, log aggregation, data transformation, and backend API services. By abstracting away infrastructure management, serverless platforms allow developers to focus solely on application logic.

Confidential serverless computing has recently attracted significant attention as it allows sensitive serverless functions to execute with hardware-backed confidentiality and isolation in untrusted cloud environments. It emerges as an important paradigm for application domains such as healthcare, finance, and privacy-preserving AI inference, where protecting sensitive data during execution is essential. In conventional serverless deployments, containers and VMs provide isolation between workloads, but the privileged cloud host and hypervisor may still access or inspect guest execution state and in-memory data. We therefore treat the cloud host and hypervisor as part of the untrusted infrastructure. Confidential Virtual Machines (CVMs) address this concern by establishing a hardware-protected execution boundary that prevents the untrusted host from directly accessing private guest memory. In the deployment model considered in this work, each serverless function is assigned a dedicated CVM, within which multiple container instances of that function may execute, while different functions are isolated in separate CVMs.

CVMs rely on hardware support such as Intel TDX~\cite{inteltdx2021}, 
AMD SEV-SNP~\cite{amd_sev_snp}, ARM CCA~\cite{armcca} to provide hardware-backed isolation and full guest-memory encryption. Achieving end-to-end confidentiality, however, requires more than the CVM execution boundary alone. CVMs are used together with mechanisms such as remote attestation and encrypted storage and communication to protect data throughout its lifecycle. The CVM mechanisms providing this protected execution environment, however, introduce non-trivial resource inefficiency and performance overhead. In particular, memory encryption disables host-level memory optimizations such as cross-VM page deduplication, while enforcing the protected execution boundary can increase the cost of events such as VMEXIT handling.

These overheads are particularly important for serverless workloads, which are typically short-lived, latency-sensitive, and dynamically scaled in response to demand. Even modest resource inefficiency or execution slowdown can therefore have a substantial impact on invocation latency, resource utilization, and operating cost. Moreover, serverless resource provisioning can interact with CVM overheads in ways that do not arise as prominently in conventional workloads. Many serverless platforms couple vCPU allocation to the configured function memory: additional CPU resources can improve latency through parallelism, but can also introduce synchronization and idle transitions when the available parallelism is not fully utilized. Such transitions may generate frequent VM exits, whose handling is substantially more expensive in CVMs. Prior work has characterized fundamental CVM overheads using general-purpose workloads~\cite{misono_confidential_2024, yan_performance_2023, qiu_price_2024}. However, it remains unclear how these overheads affect key serverless requirements, such as keeping warm containers in memory, providing low-latency warm execution, and reducing cold-start delay.

To address these gaps, we systematically examine CVM-based serverless execution along three key dimensions: memory efficiency, warm-start execution, and cold-start container initialization. For each dimension, we characterize the observed overheads, investigate their underlying causes, and identify practical mitigation opportunities.

This paper evaluates a diverse set of serverless workloads and compares their behavior across CVMs and regular VMs. Our study reveals three key findings. First, confidential deployments incur substantial memory overhead, primarily due to the loss of cross-VM page deduplication and further increased by reduced memory reclaimability. This reduces the number of warm containers that can be maintained under a fixed memory budget and can consequently increase cold starts. Second, warm-start overhead depends strongly on the workload. Functions with frequent VM exits, particularly those involving idle transitions and thread synchronization, experience substantially higher slowdown, and this effect can be amplified by serverless resource provisioning that couples vCPU allocation with memory size. Our analysis also shows the limits of existing system-level mechanisms for mitigating this behavior. Third, we characterize container creation as a major component of cold-start latency and identify bottlenecks arising from CPU configuration, cgroup contention, and container-isolation mechanisms. We further evaluate practical mitigation strategies, including, under a restricted per-function trust model, selectively relaxing container-level isolation among mutually trusted instances of the same function. Overall, these findings clarify the main performance tradeoffs introduced by confidential serverless execution and point to practical ways to mitigate them.

\section{BACKGROUND AND MOTIVATION}
\label{sec:background}

\subsection{Serverless Computing and Confidentiality Challenge}
\label{subsec:SecDemand}
Serverless workloads run in a container or a light VM on the cloud provider's infrastructure. These workloads in the cloud can suffer from side channel attacks, Denial-of-Wallet (DoW) attacks, and data being stolen by an untrusted host or hypervisor~\cite{shen_gringotts_2022, zhao2024last}.
Although containers or light VMs provide a level of isolation between different multi-tenant workloads, the hypervisor can still see the underlying memory, making users skeptical about running sensitive serverless applications inside the cloud VMs. 

One promising approach to ensure confidentiality is to run serverless functions inside Trusted Execution Environments (TEEs), such as Intel SGX enclaves \cite{intel_sgx}. These TEEs offer isolation and confidentiality guarantees by creating protected memory regions, called enclaves, that are isolated from the rest of the system; even privileged software like the Operating System or hypervisor cannot access them. This protection is mainly achieved through memory encryption and strict access control enforced by the CPU. 
Systems such as SCONE \cite{arnautov2016scone} and Graphene-SGX \cite{tsai2017graphene} demonstrate how serverless functions can be executed within enclaves to enhance data security and confidentiality in untrusted cloud environments.

Despite these benefits, the feasibility of using enclaves for serverless workloads is very limited due to several limitations. The main reason is that developers often need to significantly modify their programs to conform to enclave constraints. First, any data leaving or entering the enclave must be explicitly managed by the program to ensure confidentiality. 

Second, enclaves restrict or do not support system-level operations such as file I/O, dynamic memory allocation, and network access. This requires developers to either avoid such operations or delegate them to untrusted parts of the system \cite{priebe2020sgxlklsecuringhostos}.

\subsection{Confidential Serverless Computing: System Architecture}
\label{sec:serverlessInCvm}

To enable confidentiality, chip manufacturers provide CVM support, such as Intel TDX \cite{intel_tdx}, AMD SEV-SNP \cite{amd_sev_snp}, and ARM CCA~\cite{armcca}.
Similar to TEEs, these technologies rely on transparent memory encryption, 
typically based on AES-128 or AES-256. The encryption keys are managed by a dedicated hardware root of trust and are not accessible from the hypervisor or any software outside the CVM. Unlike traditional TEEs, CVMs extend protection to the entire VM. Thus, unmodified applications can gain 
confidentiality, provided that they execute inside a CVM. This makes CVMs particularly well-suited for serverless workloads.

The mechanisms used to provide memory encryption and isolation differ across processor architectures. For example, Intel processors rely on a trusted software component, known as the TDX Module, which runs in a CPU-protected environment and works together with hardware mechanisms to enforce protection. In contrast, AMD SEV-SNP integrates similar functionality directly within the processor, eliminating the need for a separate intermediary enforcement module. 

\begin{figure}[!t]
\centering
\includegraphics[width=1\linewidth]{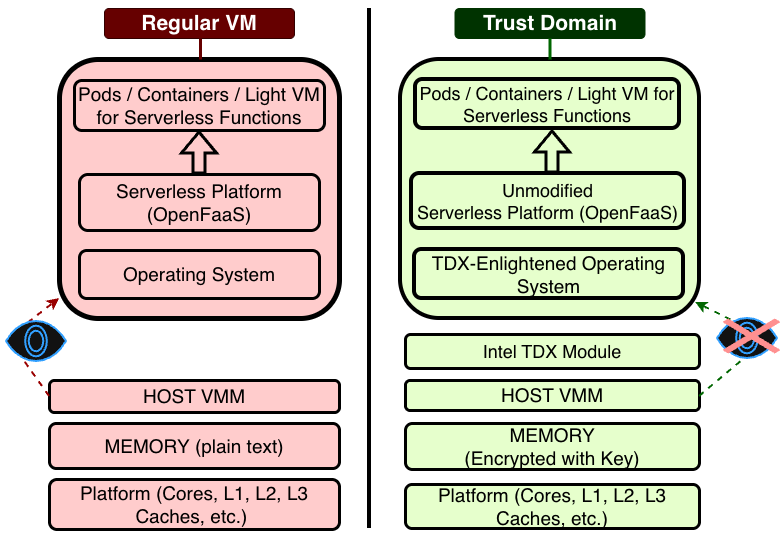}
\caption{Serverless functions on a Regular VM and Intel TD.}
\label{fig:serverless-architecture}

\end{figure}

Figure \ref{fig:serverless-architecture} compares the architectures of regular serverless computing (left) and confidential serverless computing (right), using Intel TDX as an example. In the confidential configuration, the TDX Module cooperates with hardware mechanisms to establish the VM as a protected trust domain. Inside the CVM, a TDX-aware guest OS supports confidential execution, while a serverless platform such as OpenFaaS~\cite{OpenFaas} manages functions running in containers. This allows serverless functions to execute without application modification while preventing the underlying host from directly accessing private guest memory. This protection also changes the conventional cloud trust assumption. In a regular serverless deployment, the cloud provider's privileged host software and hypervisor are trusted and may access the memory and execution state of guest VMs. In confidential computing, the host, hypervisor, and other infrastructure outside the CVM are instead considered untrusted, while the CVM hardware and software executing within the protected guest are trusted.

This hardware-protected execution boundary, however, is only one component of end-to-end confidential execution. Additional mechanisms are required to protect data outside CVM memory. Remote attestation allows a tenant to verify that a VM is authentic and is executing on genuine and trusted hardware before any keys are exchanged or any code and data are deployed. This is achieved through cryptographic measurements of the VM, typically signed by a hardware-protected key for verification. Encrypted I/O protects data transferred to and from the CVM, while encrypted storage protects data persisted outside the CVM.

\subsection{Confidential Serverless Computing: Overhead Challenges}

While the CVM architecture above provides necessary hardware-backed isolation and memory encryption to secure sensitive workloads, these protections introduce non-trivial inefficiency and performance trade-offs: the mechanisms ensuring secure trust domains encrypt guest memory also disable traditional optimization techniques; the system-level mechanisms required for isolation add complexity to management operations. Serverless functions are inherently short-lived and highly latency-sensitive. They are particularly vulnerable to even modest degradations. Particularly for the CVM architecture, they face performance challenges from two primary sources of overhead.

\subsubsection{Memory Efficiency in Serverless Deployments}
\label{sec:background-ResourceInefficienciesInCvmBasedServerlessDeployments}

The efficiency of serverless computing relies on achieving both high workload density and low latency. A major challenge is the cold start, which occurs when a function is invoked but no execution environment 
is immediately available, and thus requires the platform to create or initialize one before the function can run. This additional setup delay could be even much longer than the serverless function execution itself.

To reduce cold-start overhead, serverless platforms often keep execution environments available for reuse. An environment may be created before invocation (prewarming) or retained after a function finishes (caching), so that a subsequent invocation can reuse it with much lower latency, i.e., as a warm start. As a result, memory efficiency becomes critical, because it determines how many warm containers a cloud provider can sustain within a given memory budget while still delivering low-latency service.

In traditional virtualized environments, memory usage is often optimized through host-side deduplication mechanisms such as Kernel Same-page Merging (KSM)~\cite{arcangeli2009ksm}, which identifies and merges identical memory pages across different VMs. However, CVMs fundamentally break this optimization due to guest-memory encryption. Because each CVM uses unique hardware-managed encryption keys, identical plaintext pages, such as shared language libraries or machine learning model parameters, result in different ciphertext values, preventing the hypervisor from identifying them for merging.

CVMs can further reduce memory efficiency by limiting host-side reclamation. In regular VMs, pages that are unused or reclaimable inside the guest (e.g., file-backed pages) may still be leveraged to reduce host-observed memory footprint. However, CVMs limit this viability. 
 
The absence of cross-VM deduplication and memory reclaimability leads to substantial memory overhead. This effect is particularly pronounced in large-scale deployments and memory-intensive workloads, such as AI/ML inference. Under a fixed memory budget, the reduced memory efficiency limits the number of warm containers that can be maintained and increases the frequency of costly cold starts.

\subsubsection{Runtime Overhead}

While encryption reduces resource density, the mechanisms required to preserve the trust domain also introduce runtime overhead during function execution. This overhead may be negligible for compute-intensive serverless workloads, but it can become significant for workloads that frequently perform system-level operations. The main source of this overhead is the increased complexity of exit handling needed to maintain isolation.

In a regular VM, execution of a privileged instruction causes a VMEXIT, transferring control directly to the host VMM. The VMM can inspect the guest's register state, service the event, and resume execution with \texttt{VMRESUME}. In Intel TDX, a similar event inside a trust domain triggers a TD exit rather than a conventional VMEXIT. Instead of exposing the guest state to the VMM, the processor first transitions into the SEAM mode, where the Intel TDX Module executes. The TDX Module saves and filters the TD's architectural state, ensures that private memory remains protected, and only then forwards a sanitized exit reason to the VMM. After the VMM completes the required handling, control returns through the TDX Module using TD-specific entry instructions (e.g., \texttt{TDRESUME}), which restores the protected TD context. This additional transition path (TD $\rightarrow$ TDX Module $\rightarrow$ VMM $\rightarrow$ TDX Module $\rightarrow$ TD) introduces extra context management and isolation steps compared to regular VMEXIT handling. AMD SEV-SNP integrates memory protection mechanisms more tightly into the CPU firmware and does not introduce an intermediary enforcement module; Intel TDX explicitly separates enforcement responsibilities into the TDX Module.

\section{METHODOLOGY}

We evaluate the impact of CVMs on serverless computing in terms of memory efficiency and runtime performance by conducting experiments on real systems using a diverse set of workloads.

\subsection{Experiment Design}

To isolate the cost of confidentiality itself, we compare the serverless deployments in confidential and regular VMs on both cloud platforms and a local testbed. The cloud platforms provide diversity in hardware architectures and configurations, including Intel TDX and AMD SEV-SNP. This helps avoid conclusions from a single confidential computing implementation. The local testbed enables finer-grained profiling and root-cause analysis that is difficult to perform in public clouds.

We run a diverse set of serverless functions for broad workload coverage. To ensure reliable and consistent measurements, we minimize sources of runtime variability and keep the experimental environment stable. Specifically, we maintain a consistent software stack across all configurations. We also reduce variability from function scheduling and network behavior. To this end, we directly operate functions using pre-built containers. Container images and function artifacts, such as models and input files, are provided locally so that results are not affected by external network bandwidth fluctuations. Experiment setup and workload details are in Section~\ref{sec:exp_setup}.

\subsection{Memory-Efficiency Evaluation}
\label{subsection:evalMemoryEfficiencyinCVM}

To quantify the memory cost, we compare the host-observed memory usage of serverless deployments under three configurations: regular VMs with KSM enabled, regular VMs with KSM disabled, and CVMs. The comparison between KSM-enabled and KSM-disabled regular VMs measures the memory cost caused 
by the loss of cross-VM deduplication over actively used identical pages. The comparison between KSM-disabled regular VMs and CVMs identifies additional memory inefficiency introduced by reduced memory reclaimability.

Because this overhead may vary with deployment scale, we study two scaling dimensions: the number of VMs and the number of container instances per VM. As more functions are deployed and more memory resources are required to support them, the memory overhead can be in turn amplified. 

We focus on the memory consumed by cached execution environments, namely warm containers and the associated runtimes, rather than the transient memory used during function execution. There are a few reasons: 1) execution-time memory usage is highly dynamic and varies over time. This leads to unstable measurements; 2) due to its transient nature, execution-time memory is rarely deduplicated in current practice; systems also typically reserve sufficient memory for execution; 3) cached environments typically account for the dominant share of memory consumption~\cite{yu_rainbowcake_2024}.

\subsection{Runtime-Overhead Evaluation}

\begin{figure}[!t]
\centering
\includegraphics[width=1\linewidth]{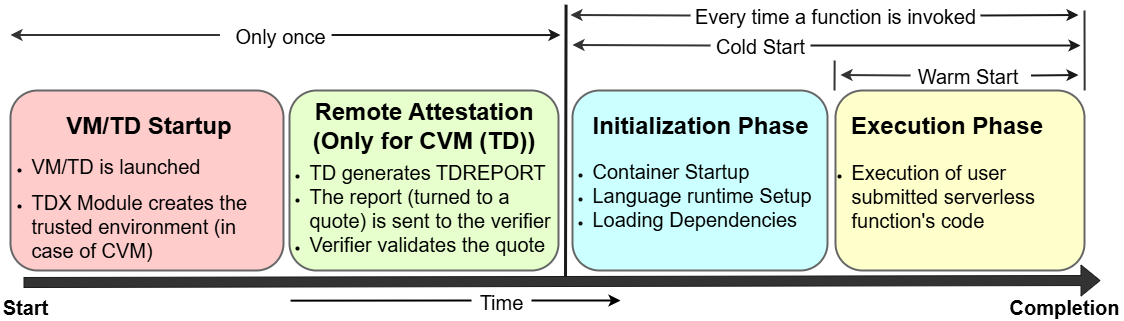}

\caption{Execution of a Serverless Function.}

\label{fig:serverlessExec}

\end{figure}

While a systematic evaluation of runtime overhead should cover the entire serverless lifecycle, it is neither meaningful to treat the whole lifecycle as a single end-to-end quantity nor necessary to examine every stage in equal depth. To maintain a systematic view while focusing on the most meaningful overheads, we divide the serverless function lifecycle into four phases (Figure~\ref{fig:serverlessExec}) and consider them 
separately. For the phases most relevant to user-perceived serverless performance, we further compare their behavior on regular VMs and CVMs.

\noindent$\bullet$ \textbf{VM/TD Startup phase} launches the VM or trust domain (TD) that will host the serverless functions. Prior work~\cite{misono_confidential_2024} has studied this phase and reports that VM boot latency increases by up to 231\% for AMD SEV-SNP and up to 394\% for Intel TDX CVMs compared to regular VMs.

\noindent$\bullet$ \textbf{Remote Attestation phase} verifies that the VM is authentic and running on trusted hardware before provisioning secrets or sensitive data. Its overhead has also been quantified in prior work and is reported to be marginal, typically under 10 ms~\cite{misono_confidential_2024}. 

\noindent$\bullet$ \textbf{Initialization phase} prepares the execution environment for a function, including container creation, runtime loading, and dependency setup (libraries, key datasets, etc). It is the main source of cold-start overhead when a warm execution environment is not already 
available.

\noindent$\bullet$ \textbf{Execution phase} covers the actual execution of the function once the execution environment has been prepared (i.e., warm starts). The time is determined by the function logic and affected by CVM runtime overhead.

Not every phase needs to be examined in equal detail in this work. VM startup and remote attestation have already been measured in prior studies and are incurred only in selected scenarios. In contrast, initialization and execution more directly affect the latencies observed by users in warm-start and cold-start scenarios. The overhead incurred in these phases has not been studied. Therefore, our evaluation focuses on these phases when comparing regular VMs and CVMs.

\subsection{Root-Cause Analysis Strategy}

When slowdowns under confidential execution are identified, we perform a root-cause analysis to determine the root cause. This analysis is conducted on the local testbed with controlled experiments. We first profiled the system and hardware counters for both CVM and regular VM executions to identify possible contributing factors. 
We then control the experimental setting to identify the underlying cause of the contributing factor.

\section{EXPERIMENTAL SETUP}
\label{sec:exp_setup}

\subsection{Evaluation Platform}
\label{subsection:evalPlatform}

We evaluate our system on Google Cloud Platform (GCP) using Intel Sapphire Rapids servers with TDX support and AMD Milan servers with SEV-SNP support. On the Intel platform, we use \texttt{c3-standard-22} instances with 22 vCPUs and 88 GB of memory, and c3-standard-4 instances with 4 vCPUs and 16 GB of memory. On the AMD platform, we use \texttt{n2d-standard-4} instances, which match the smaller Intel configuration with 4 vCPUs and 16 GB of memory. For each instance type, we evaluate both confidential VMs (CVMs) and non-confidential VMs (NVMs). All cloud instances use GCP standard persistent disks with Google-managed encryption. We also conduct selected experiments on a local \textit{HPE ProLiant DL380 Gen11} server with dual-socket \textit{Intel Xeon Gold 6554S} CPUs (144 logical cores in total). All VMs and the local host run Ubuntu Linux with kernel version 6.8.0. Additional system details are summarized in Table~\ref{table: System configurations}.

\begin{table}[!t]
\caption{System configurations of Intel and AMD Machines\label{table: System configurations}}
\vspace{-0.05in}
\centering
\begin{tabular}{|p{0.24cm}|p{1.7cm}|p{0.50cm}|p{1.8cm}|p{0.90cm}|p{0.46cm}|} 
\hline
SN & Processor & CVM & Machine Type & \# vCPU (Core) & Mem. (GB) \\
\hline
    1 & Intel Sapphire & Yes & c3-standard-4 & 4 (2) & 16 \\ 
\hline
    2 & Intel Sapphire & No & c3-standard-4 & 4 (2) & 16 \\ 
\hline
    3 & Intel Sapphire & Yes & c3-standard-22 & 22 (11) & 88 \\ 
\hline
    4 & Intel Sapphire & No & c3-standard-22 & 22 (11) & 88 \\ 
\hline
    5 & Intel Xeon & Yes & Local machine & 32 (32) & 16 \\ 
\hline
    6 & Intel Xeon & No & Local machine & 32 (32) & 16 \\
\hline
    7 & AMD Milan & Yes & n2d-standard-4 & 4 (2) & 16 \\ 
\hline
    8 & AMD Milan & No & n2d-standard-4 & 4 (2) & 16 \\
\hline
\end{tabular}
\vspace{-0.2in}
\end{table}

\subsection{Serverless Platform and Deployment Model}

We use OpenFaaS \cite{OpenFaas} as the serverless computing framework in our setup, deploying it as a single-node instance. OpenFaaS is orchestrated using Kubernetes, which handles the scheduling and management of all associated pods. All serverless function pods, including system-level pods such as the OpenFaaS gateway, queue worker, and Prometheus, are deployed on the same machine to minimize network variability and focus on the underlying VM and hardware impacts.

This setup ensures that function scheduling, execution, and metrics collection are localized, allowing for fine-grained measurement of performance differences between configurations. We ensure the function containers are prebuilt and ready to be instantiated for warm start, isolating function runtime behavior from image pulling or container setup delays. For cold starts, we ran the containers from the start.

\subsection{Workloads and Benchmarks}
\label{sec:WorkloadsAndBenchmarks}

We selected a diverse set of serverless functions from SeBS~\cite{copik_sebs_2020}, FunctionBench \cite{kim_functionbench_2019}, and FaaSProfiler \cite{shahrad_architectural_2019} benchmark suites to cover a range of application domains and workload characteristics (e.g., compute-intensive and I/O-intensive) functions. Table~\ref{table:benchmark_lists} summarizes the benchmarks, which can be grouped into four categories:

\noindent$\bullet$ \textbf{Compute-Intensive:} arithmetic-heavy and CPU-bound functions, including \textit{Matrix Multiplication, Map Reduce, Video Processing}, and \textit{Graph BFS}.

\noindent$\bullet$ \textbf{I/O-Intensive:} functions involving significant disk or file operations, including \textit{Compression}, and \textit{Random Disk I/O}.

\noindent$\bullet$ \textbf{ML \& Inference Workloads:} machine learning inference functions or analysis tasks, including \textit{Image Recognition (IR), Video Face Detection, Logistic Regression, RNN Generate Character Level, CNN Image Classification}, and \textit{Sentiments}.

\noindent$\bullet$ \textbf{Web, Encryption, API, and Backend Logic:} application-oriented serverless functions, including \textit{Chameleon, CRUD API}, and \textit{Pyaes}.

\begin{table}[!t]
\centering
\begin{threeparttable}

\captionsetup{justification=centering, singlelinecheck=false}
\caption{List of different benchmarks used \\
\small Note:FB---FunctionBench, FP---FaaSProfiler
} 
\label{table:benchmark_lists}
\begin{tabular}{|p{0.16cm}|p{2.44cm}|p{0.6cm}|p{3.87cm}|} 
\hline
SN & Benchmarks & Source & Task \\
\hline
    1 & Chameleon & FB & Evaluates template rendering cost\\
\hline
    2 & Matrix Multiplication & FB & Multiplication of 2 nxn matrices\\
\hline 
    3 & Compression & SeBS & Compresses files  in a directory\\ 
\hline
    4 & Video Processing & SeBS & GIF generation \& watermark insertion\\ 
\hline
    5 & Image Recognition & SeBS & Inference on an image\\ 
\hline
    6 & Pyaes & FB & Performs AES encryption\\
\hline
    7 & Video Face Detection & FB & Face detection frame by frame\\
\hline
    8 & Map Reduce & FB & MapReduce pipeline\\
\hline
    9 & Crud API & SeBS & Create-Read-Update-Delete\\
\hline
    10 & Random Disk IO & FB & Random reads writes to disk\\
\hline
    11 & RNN Generate Character Level & FB & Generate names\\
\hline
    12 & CNN Image Classification & FB & Identifies an image\\ 
\hline
    13 & Logistic Regression & FB & Training a model\\ 
\hline
    14 & Graph BFS & SeBS & Breadth first search on a graph\\
\hline
    15 & Sentiments & FP & Parsed and analyzed text\\ 

\hline
\end{tabular}

\end{threeparttable}
\end{table}

Some of the functions listed above may fall under multiple categories. For clarity, however, each function is assigned to only one category in the analysis. All functions are deployed using OpenFaaS templates and invoked through HTTP requests. We keep resource limits and environment variables consistent across all deployments to ensure fair comparison across regular VM and CVM configurations.
\section{EXPERIMENTS AND FINDINGS}
\label{sec:expAndFindings}

This section first evaluates how CVM affects memory efficiency (Section~\ref{subsec:mem_efficiency}), since the loss of cross-VM page deduplication and memory reclaimability directly impacts resource provisioning and deployment density. We then analyze the execution overhead of these benchmarks under warm-start conditions (Section~\ref{subsec:perf_overhead}), followed by cold-start behavior (Section~\ref{subsec:cold_start}).

\subsection{Memory Efficiency} 
\label{subsec:mem_efficiency}
Unlike prior CVM studies that mainly emphasize execution overhead on general-purpose workloads, our analysis explicitly quantifies how confidentiality changes memory scaling behavior in serverless systems, where warm-container capacity is a first-order performance concern. Our study aims to clarify three questions: (i) Does confidentiality introduce memory cost primarily because it disables cross-VM deduplication? (ii) How is this cost affected by factors such as deployment scale and workloads? (iii) What does that imply for warm-container capacity in serverless systems? 

To answer these questions, we compare three configurations: regular VMs with KSM enabled, regular VMs with KSM disabled, and CVM. The comparison between KSM-enabled and KSM-disabled regular VMs isolates the cost of losing cross-VM deduplication, while the comparison between KSM-disabled VMs and CVMs identifies the additional inefficiency caused by reduced memory reclaimability. To ensure realistic measurements, we provision each VM with the minimum memory required for correct and stable function execution. This avoids attributing excess unused memory to the system footprint, which would otherwise inflate the measured memory consumption and overstate the resulting overhead. Although some degree of memory overprovisioning is common in cloud practice, there is no fixed standard for how much extra capacity should be reserved. Using a tight memory bound, therefore, provides a more controlled and interpretable basis for comparison.

\subsubsection{\textbf{VM instances for same functions}}
\begin{figure*}[!t]
\centering
\includegraphics[width=1\textwidth]{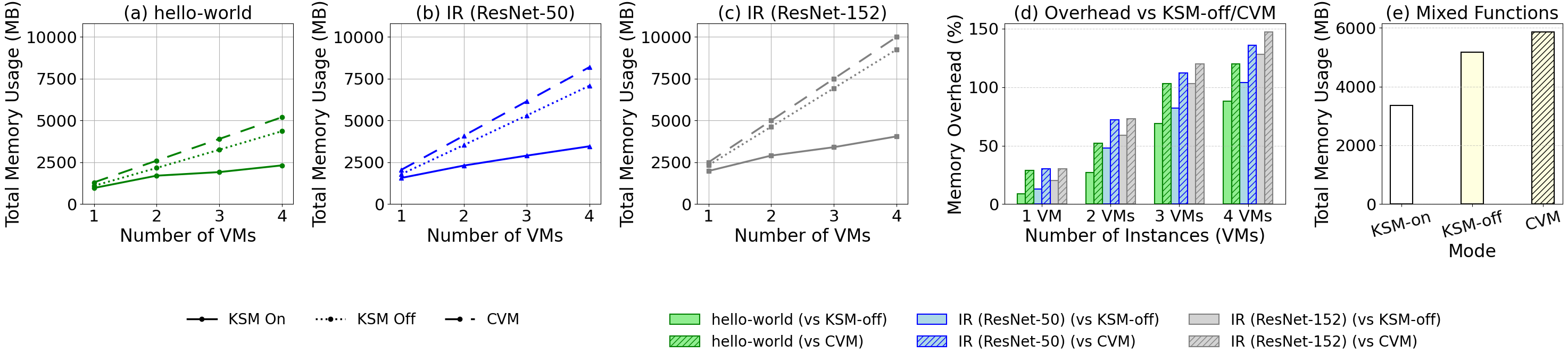}
\caption{Memory usage across different numbers of VMs for functions (a) hello-world, (b) IR (ResNet-50), and (c) IR (ResNet-152); (d) Memory inefficiency caused by disabling page deduplication across serverless workloads; (e) Total memory usage when running different functions}
\label{fig:ksm_memory_all}
\end{figure*}

We first examine the VMs created for hosting the same function, one instance in each VM. Figures~\ref{fig:ksm_memory_all}(a)–(c) show total memory usage as the number of VMs increases for three representative functions with memory demands ranging from lightweight ({\textit{hello-world}}) to memory-intensive (IR (\textit{ResNet 152}) --- Image recognition using ResNet 152). Across all workloads, memory usage in both the KSM-disabled and CVM configurations grows much faster than in KSM-enabled executions. 

\paragraph{Scaling with Number of VMs}
For lightweight workloads such as hello-world (Figure~\ref{fig:ksm_memory_all}(a)), total memory increases from about 1 GB to 2.3 GB across 1–4 VMs with KSM, compared with 4.4 GB and 5.2 GB for KSM-disabled and CVM configurations, respectively.

Similar trends are also observed for IR. With ResNet-50 (Figure~\ref{fig:ksm_memory_all}(b)), memory usage under KSM-enabled execution increases from roughly 1.5 GB to 3.4 GB, whereas KSM-disabled and CVM configurations reach approximately 7.1 GB and 8 GB at 4 VMs. With ResNet-152 (Figure~\ref{fig:ksm_memory_all}(c)), the gap is even larger: KSM-enabled memory grows from about 2 GB to 4 GB, while KSM-disabled and CVM configurations reach nearly 9.2 GB and 10 GB, respectively.

These results show that, without deduplication, each VM contributes nearly the full memory footprint of the workload, while KSM-enabled execution benefits from cross-VM page sharing. The close alignment between CVM and KSM-disabled curves confirms that the loss of deduplication is the dominant source of memory inefficiency in CVMs.

\paragraph{Impact of Workload Size}
Figure~\ref{fig:ksm_memory_all}(d) shows how the memory overhead varies with workload size. Even for a single VM, the absence of deduplication introduces 9–20\% overhead, which increases significantly with scale. At four VMs, the overhead reaches approximately 88\% for hello-world, 104\% for ResNet-50, and 128\% for ResNet-152 under KSM-disabled execution. For CVMs, the overhead is even higher, reaching approximately 120–147\% for the largest configurations.

These results show that memory inefficiency grows with workload size. Larger models contain more identical memory pages (e.g., model weights, runtime components, and shared libraries), which would otherwise be consolidated in KSM-enabled environments. At the same time, a larger memory space is reserved for their execution. As a result, the absence of memory deduplication and reclamation leads to higher overhead for memory-intensive workloads.

\subsubsection{\textbf{VM instances for different functions}}

To show that the cost of confidentiality is not limited to homogeneous deployments with repeated copies of the same workload, we create three VMs hosting three different workloads (hello-world, ResNet-50, and ResNet-152), and measure their total memory consumption under the above three configurations. Figure~\ref{fig:ksm_memory_all}(e) shows that memory inefficiency persists even when the VMs host different functions. Under KSM-enabled execution, substantial memory savings are still possible because VMs share system-level pages such as OS components, runtime environments, and common libraries. Prior work \cite{medes_dedup} also confirms this claim by stating that runtime and libraries that are common across the functions determine the extent of redundancy different functions' sandboxes have. Under CVMs, these optimizations are not available, and memory usage again closely follows the KSM-disabled configuration. Thus, the loss of deduplication under CVMs is a platform-wide scaling issue, not merely an artifact of identical-function replication.

\subsubsection{\textbf{Implications for Warm-Container Capacity}}
\label{subsubsection:implicationsForWarmContainerCapacity}

The results suggest that the dominant memory penalty in CVM-based serverless deployment is from the loss of cross-VM page deduplication. In regular VM deployments, consolidating identical pages across VMs effectively increases usable memory capacity under a fixed memory budget. Confidential deployments do not benefit from this effect.

This difference directly affects the capacity of caching warm containers. In regular deployments, reclaimed memory can be used to maintain a larger warm container pool. This helps reduce cold starts. In CVM-based deployments, the same memory budget supports fewer resident warm instances, which can lead to increased cold starts.

This then motivates a targeted question: although confidentiality prevents cross-VM deduplication, can memory-sharing opportunities be exploited within a single VM to recover the lost efficiency through intra-VM memory deduplication? To answer this question, we compare two deployments with the same total number of container instances: 4VM-1Instance, in which four VMs each host one container instance, and 
1VM–4Instances, in which one VM hosts four container instances. The results confirm that consolidating four container instances within a single VM substantially 
reduces total memory usage compared with distributing the same four instances across four separate VMs. For the hello-world workload, memory usage decreases from 4.3 GB (KSM disabled) and 2.3 GB (KSM enabled) in the 4VM setup to 1.1 GB when all instances are colocated in a single VM. Similarly, for IR (ResNet-50), memory reduces from 6.9 GB / 3.4 GB to 2.2 GB, and for IR (ResNet-152), from 9.0 GB / 4.0 GB to 2.9 GB. 

From a systems perspective, these results suggest that hosting multiple instances of the same function within a single CVM can mitigate the memory inefficiency of confidential deployment and improve warm-container capacity. This opportunity is particularly relevant because serverless platforms frequently experience concurrent requests to the same function, requiring multiple instances of that function to coexist~\cite{concurrent_asplos25}.

\subsection{Overheads Analysis in Warm Starts}
\label{subsec:perf_overhead}

This subsection focuses on performance overhead under the most common operating condition—warm starts. We first evaluate warm-start execution latency of benchmark functions on CVMs and then analyze the root causes of the largest slowdowns.

\begin{figure}[!t]
\centering
\includegraphics[width=1\linewidth]{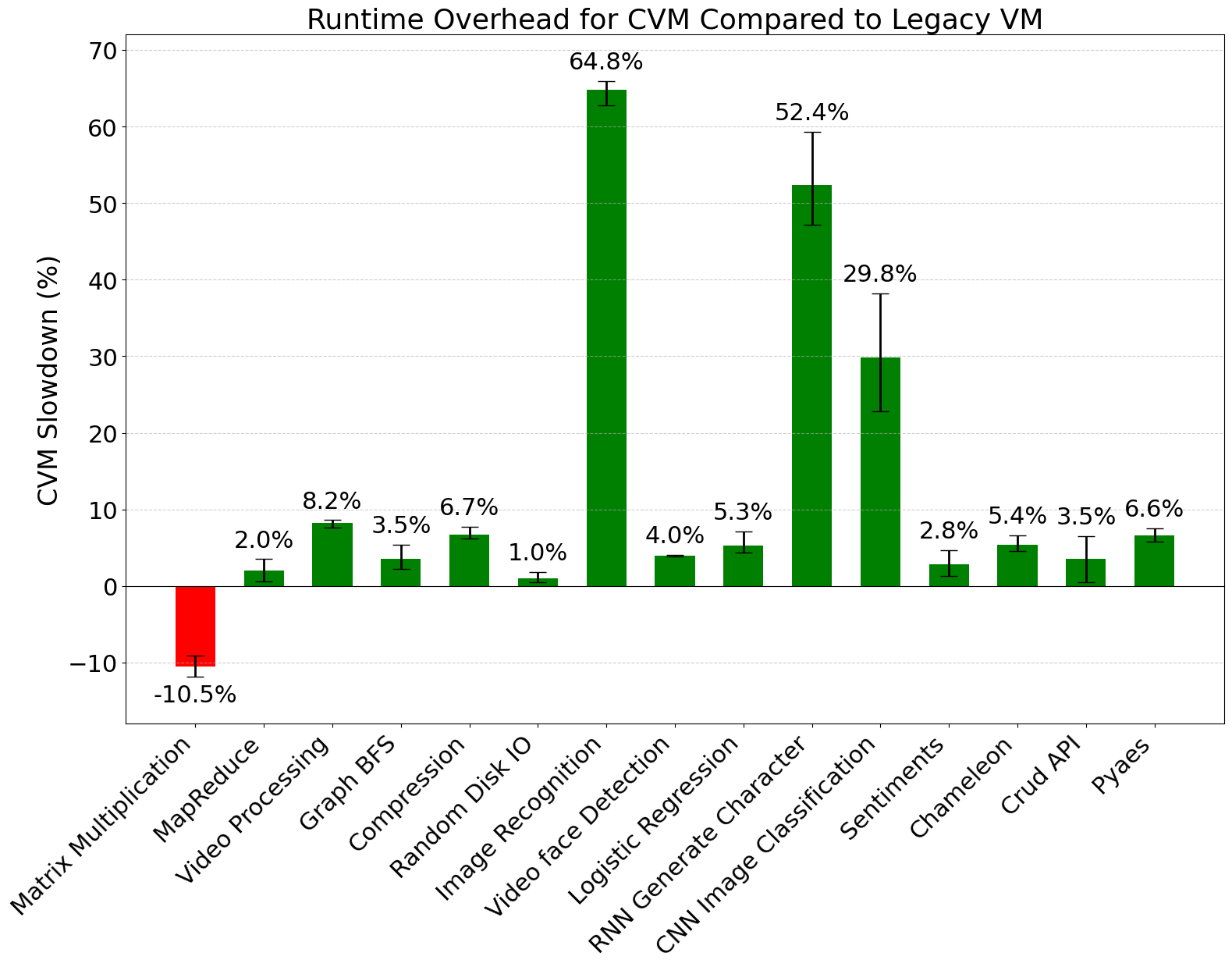}
\caption{Latency per invocation across different serverless workloads on Intel TDX machines.}
\label{fig:latency_invocation}
\end{figure}

We evaluate the warm-start latencies for the benchmarks listed in Section~\ref{sec:WorkloadsAndBenchmarks} on both regular VMs and CVMs. Figure~\ref{fig:latency_invocation} presents the latency overhead we find in CVM. The results show that the performance impact of confidential execution is strongly workload dependent. Several compute-intensive and I/O-oriented workloads exhibit only minor slowdowns under CVMs, typically within a single-digit percentage range. This indicates that workloads dominated by straightforward CPU computation or disk operations remain largely unaffected by confidential execution.

In contrast, certain workloads experience substantial performance degradation. In particular, several machine learning inference benchmarks—including Image Recognition, RNN Character Generation, and CNN Image Classification—exhibit noticeable slowdowns under CVMs, with Image Recognition showing the largest degradation among all workloads.
Overall, these findings indicate that warm-start overhead in CVMs is concentrated in specific workload behaviors rather than uniformly distributed across workloads. We next analyze the underlying causes of these slowdowns.

\subsubsection{\textbf{VMEXIT-Induced Serverless Execution Overheads}}

\begin{figure}[!t]
\centering
\includegraphics[width=0.85\linewidth]{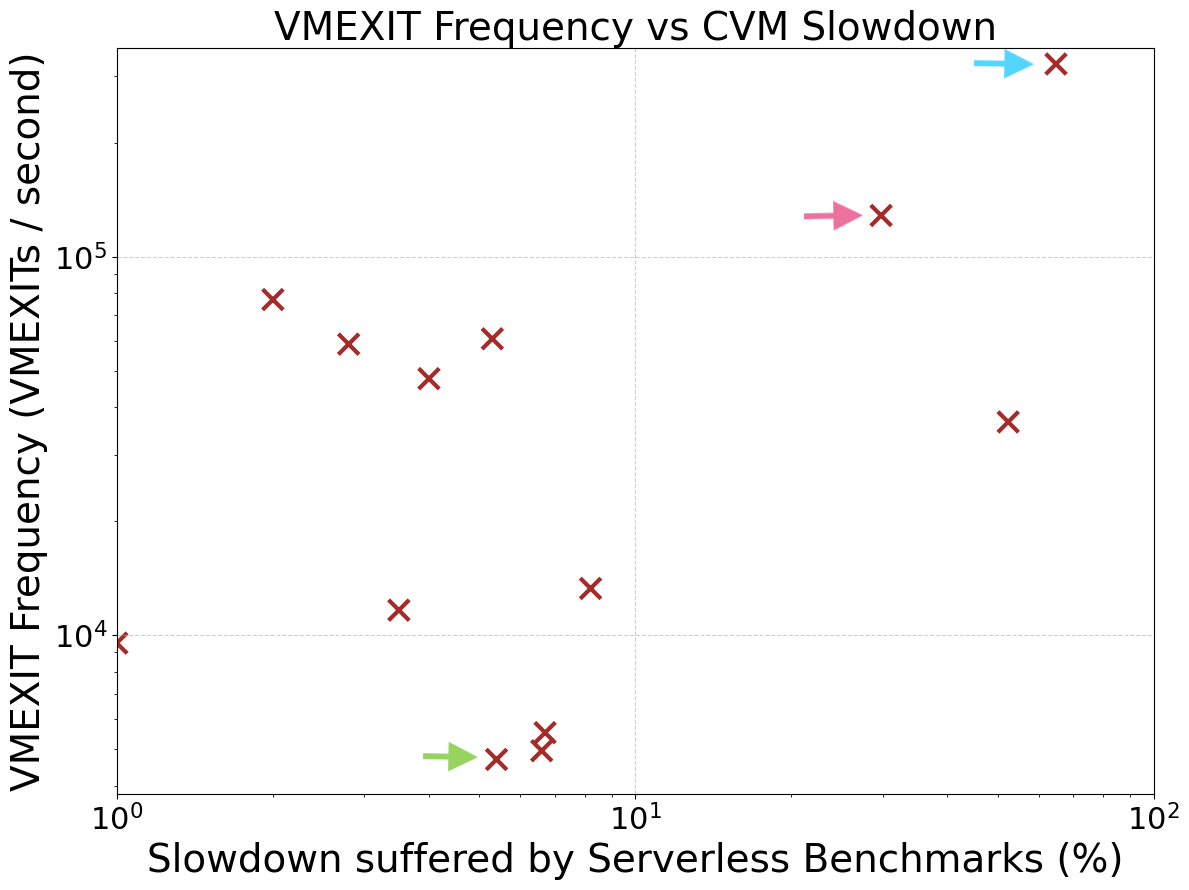}
\caption{Frequency of VMEXITs for different serverless functions based on the slowdown they suffered}
\label{fig:vmexits_frequency}
\end{figure}

Prior work~\cite{misono_confidential_2024} has identified VMEXITs as a major source of overhead for general-purpose workloads in CVMs. To investigate the source of warm-start slowdowns for serverless functions, we first examine the relationship between execution slowdown and VMEXIT activity across the workloads. Figure~\ref{fig:vmexits_frequency} shows that workloads with larger slowdowns generally exhibit higher VMEXIT rates. For example, Image Recognition (IR) shows the largest slowdown; it also shows the largest (very high) VMEXIT rate, more than 300 thousand per second (shown by the blue arrow in Figure~\ref{fig:vmexits_frequency}). CNN Image Classification generates fewer VMEXITs than IR (around 130 thousand per second, shown by the red arrow in Figure~\ref{fig:vmexits_frequency}) and exhibits a lower slowdown. Chameleon generates only a few thousand VMEXITs each second (shown by the green arrow in Figure~\ref{fig:vmexits_frequency}) and thus experiences minimal degradation.

VMEXIT frequency alone does not fully explain performance degradation. Some benchmarks, such as Video Face Detection and Logistic Regression, exhibit relatively high VMEXIT frequency but do not experience significant slowdown. This is because a large fraction of their VMEXITs are low-cost types, such as \texttt{EXT\_INTERRUPT} and \texttt{MSR\_WRITE}. In contrast, workloads that generate a higher proportion of expensive VMEXITs—particularly \texttt{HLT}, which is associated with idle transitions—experience significantly higher slowdown. This aligns with prior findings~\cite{misono_confidential_2024}, which identify that vCPU-sleep (HLT) overhead of the CVM (may be incurred by frequently repeating sleep 
and wake-up events) can cause a significant performance drop and suggests that enforcing guest-side polling can help mitigate this issue.

\subsubsection{\textbf{Impact of Parallelism on Execution Overheads}}

To understand how this known CVM overhead manifests in serverless warm starts, particularly its interaction with serverless deployment practice, we examine the role of runtime-managed multithreading. In many serverless platforms, assigning more memory to a function is commonly associated with allocating more vCPUs. Runtime frameworks (such as PyTorch, Intel TBB) often respond by spawning worker threads based on the number of vCPUs. For workloads with frequent synchronization, using an extra number of threads creates more idle–active transitions in vCPUs and therefore more \texttt{HLT}-triggered exits and amplified CVM overhead.

\begin{figure}[!t]
\centering
\includegraphics[width=0.75\linewidth]{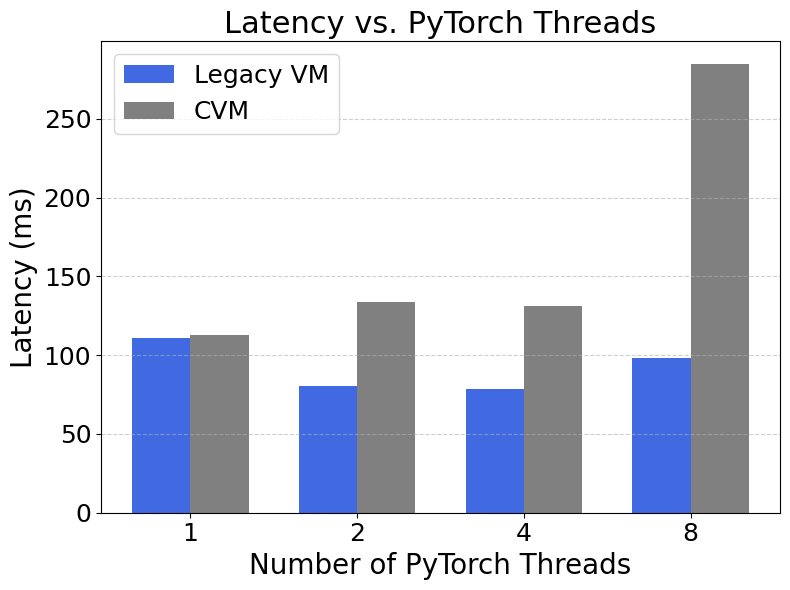}
\caption{Inference time for Image Recognition when using different PyTorch threads on a 2-core Intel TDX Machine.}
\label{fig:latencies_multi_threaded}
\end{figure}

To quantify this effect, we control the number of vCPUs and run the IR benchmark. PyTorch used by the IR benchmark automatically adjusts the number of threads based on the number of vCPUs. As shown in Figure~\ref{fig:latencies_multi_threaded}, overall, single-thread latencies are similar between CVMs and regular VMs, whereas multi-thread latencies are substantially higher on CVMs. On a regular VM, increasing the thread count generally reduces latency. In contrast, on a CVM, single-threaded execution outperforms multi-threaded execution. This behavior is consistent with additional synchronization and idle transitions generated during multithreaded execution, whose associated VMEXIT costs are amplified under CVMs.

The overhead associated with idle-transition VMEXITs is not unique to confidential execution and has also been observed in regular VMs~\cite{gleaner}. However, the effect is considerably more severe in CVMs. For example, at eight threads, both environments suffer from synchronization overhead, but the slowdown is much larger in the CVM, reaching nearly 200\% (Figure~\ref{fig:latencies_multi_threaded}). 
To further highlight this effect, we repeat the experiment on larger VMs with 22 vCPUs. The slowdown increases to 3.5× for executions on CVM, while the executions on the regular VM can still show a small latency improvement over the single-thread execution.

\begin{table}
  \caption{IR inference latency and VMEXIT activity under different guest idling policies. \label{table:latenciesVsIdling}}
  \centering
\begin{tabular}{|c|c|c|c|}
\hline
    VM Type & Idling Policy & Avg. Latency (ms) & \# VMEXITs \\
    \hline
    Regular & HLT & 201.4 & 3.6M \\ 
    \hline
    CVM & HLT & 347.8 & 4.5M \\ 
    \hline
    Regular & POLL & 30.8 & 536K \\ 
    \hline
    CVM & POLL & 31.8 & 679K \\ 
\hline
\end{tabular}
\end{table}

To isolate the contribution of idle-transition VMEXITs, we perform a controlled ablation by changing only the guest idling policy from the default HLT policy to POLL\footnotemark. Table~\ref{table:latenciesVsIdling} shows both inference latency and VMEXIT activity. Under HLT, the regular VM and CVM incur average latencies of 201.4 ms and 347.8 ms, respectively, together with substantially higher VMEXIT counts. Switching to POLL reduces the corresponding average latencies to 30.8 ms and 31.8 ms, while total VMEXIT counts decrease from 3.6 millions to 536 thousands for the regular VM and from 4.5 millions to 679 thousands for the CVM. Notably, under the HLT policy in the regular VM, HLT accounts for 29\% of the total VMEXITs, but contributes approximately 99\% of the cumulative time attributed to VMEXIT handling. Under POLL, HLT exits are eliminated entirely. Thus, suppressing idle transitions substantially reduces both VMEXIT activity and execution latency under otherwise unchanged execution conditions. This controlled result provides stronger evidence that idle-transition VMEXITs are a major contributor to the observed slowdown.

\footnotetext{
An operating system can employ different idling policies to determine how a CPU behaves when it becomes idle. Under the \texttt{HLT} (halt) policy, a CPU calls the HLT instruction to put itself into a low-power mode. In the case of VMs, a vCPU calling the HLT instruction will cause a VMEXIT, which returns control to the hypervisor and may in turn trigger a context switch to another vCPU or task on the host. In contrast, the \texttt{POLL} policy keeps the vCPU active by having the vCPU call the \texttt{PAUSE} instruction in a loop (i.e., \texttt{PAUSE}-loop), thereby removing the VMEXIT caused by HLT instructions.}

\subsubsection{\textbf{Existing system-level mitigations and their limitations}}

A straightforward way to reduce costly idle transitions is to keep vCPUs spinning rather than allowing them to enter an idle state, as demonstrated by our POLL experiment above. However, spinning introduces its own cost by consuming CPU cycles while no useful work is being performed. This creates a fundamental trade-off: a short spinning period may fail to avoid costly idle transitions, whereas a long spinning period can waste CPU resources and interfere with co-located workloads. Existing system-level mechanisms therefore attempt to balance these two costs by dynamically adjusting the duration of spinning. Prior work has also explored more fundamental ways to reduce spinning overhead, such as consolidating fine-grained tasks onto fewer vCPUs to shorten idle periods~\cite{gleaner} and shifting waiting activity to SMT sibling cores~\cite{vSMTIO}. Nevertheless, these approaches do not eliminate the underlying trade-off, since their effectiveness still depends on adapting to the workload's idle behavior while balancing idle-transition overhead against unnecessary CPU consumption.

Host-level spinning provides one way to reduce idle-transition overhead. Linux/KVM implements adaptive halt polling (\texttt{halt\_polling}), in which a vCPU briefly polls at the host before being descheduled. This mechanism is enabled by default for both regular VMs and CVMs. However, substantial slowdown remains for the serverless workloads we evaluate. The ineffectiveness may be partially explained by serverless functions being short-lived and their execution phases changing too quickly for halt polling to adapt. Also, host-level spinning mechanisms cannot eliminate VMEXITs associated with guest-level idle transitions. 

Guest-level polling avoids these HLT-triggered exits, as shown in our previous experiment, but continuously spinning vCPUs can waste CPU resources and interfere with co-located workloads. In regular VMs, this overhead can be controlled using the \texttt{PAUSE-Loop Exit} (PLE): when a guest repeatedly executes the \texttt{PAUSE\_INSTRUCTION} instruction, the hypervisor can detect the spin loop and preempt the vCPU, improving scheduling fairness and reducing wasted CPU cycles~\cite{APPLES}. CVMs, such as Intel TDX, however, do not expose the same hardware-assisted PLE mechanism and instead rely on approaches such as paravirtualized spinlocks, which require hypervisor cooperation and may be undesirable when the hypervisor is untrusted. This makes efficient control of guest-side spinning more difficult in CVMs.

For CVM, thus the best possible system-level approach would be consolidating threads onto vCPUs to reduce
idle periods and then using short polling to reduce resource waste, a similar approach as Gleaner~\cite{gleaner}.
However, this can significantly increase latencies, and thus cannot be used for serverless computing. 

To evaluate this trade-off, we run two regular VMs (configuration 6 in Table~\ref{table: System configurations}) pinned to the same physical CPU cores. One VM executes five concurrent Image Recognition instances to reduce idle periods, while the second runs the memtier benchmark~\cite{memtier_benchmark}. 
We evaluated two configurations: (i) PLE enabled: The default configuration, allowing the hypervisor to preempt spinning VMs via the \texttt{PAUSE\_INSTRUCTION}, and (ii) PLE disabled: This configuration emulates CVM behavior, where the PLE is not observed. We use two regular VMs rather than a regular VM and a CVM to isolate the effect of PLE from other CVM-specific performance differences.

\begin{figure}[!t]
\centering
\includegraphics[width=0.85\linewidth]{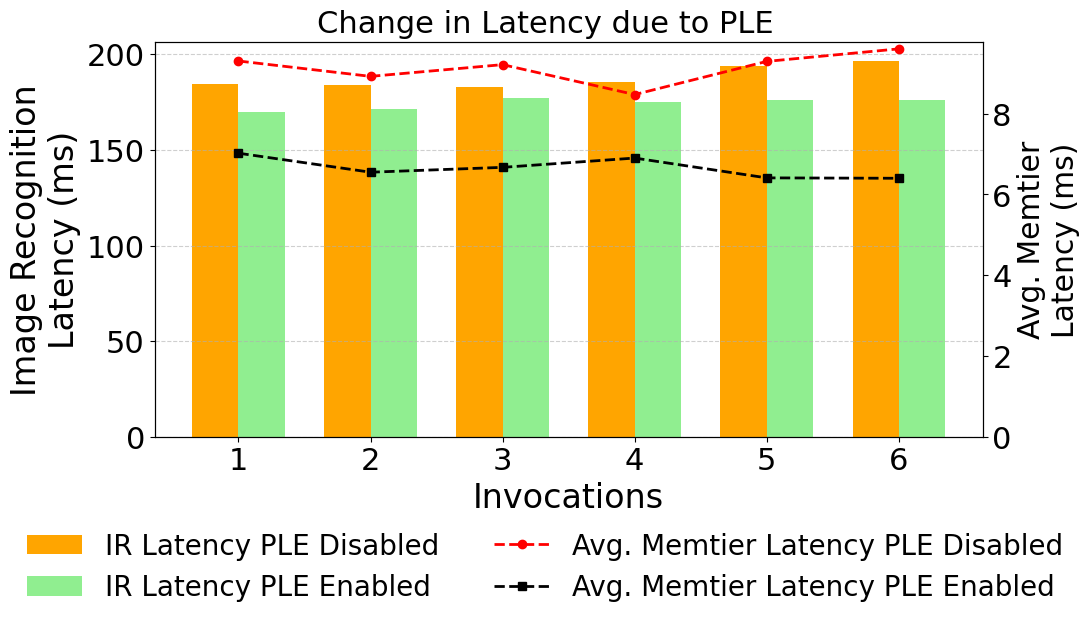}
\caption{Latency of serverless and memtier benchmarks, enabling and disabling the PLE.}
\label{fig:pleLatency}
\end{figure}

Figure~\ref{fig:pleLatency} shows that the latencies remain higher than those observed without workload consolidation. To further clarify, the latency increase was not due to aggressive consolidation. Although five concurrent IR instances keep the vCPUs busier, short idle intervals still occur between computation phases. Under guest-side polling, these remaining idle intervals cause vCPUs to spin, which can interfere with co-located workloads. To quantify the importance of controlling this spinning, we compare PLE-enabled and PLE-disabled configurations. Disabling PLE increases latency for both workloads: IR experiences an 8\% slowdown, while memtier degrades by 37\%.

These limitations motivate controlling parallelism closer to the function runtime. Rather than reacting to idle transitions after they occur, a serverless platform could use execution history to select an appropriate concurrency level (e.g., thread count) for each function, reducing unnecessary synchronization and idle-active transitions in the first place.

\subsubsection{\textbf{Other System-Level Effects}}While the previous analysis explains how VMEXIT frequency, thread-level parallelism, and idling behavior contribute to performance degradation for serverless functions, in this section, we first establish generality, then analyze some other system-level mechanisms that we observed.

\paragraph{Cross-Platform Validation (AMD SEV-SNP)} To determine whether the observed slowdown is specific to Intel TDX or characteristic of CVMs more broadly, we repeated the IR benchmark on two-core AMD SEV-SNP instances (configurations 7 and 8 from Table~\ref{table: System configurations}). Similar to the Intel platform, the serverless benchmark on AMD SEV-SNP also faces a performance slowdown of around 10\% as compared to a non-confidential regular AMD VM. Although the degradation on AMD is less severe than on Intel TDX, the presence of comparable behavior indicates that this slowdown is not unique to Intel TDX and can also appear on another confidential computing implementation.

\paragraph{Hardware Placement Effects}
While the slowdown behavior is consistent across confidential computing implementations, the unexpected performance improvement observed for matrix multiplication requires separate investigation.
To investigate whether this behavior arises from hardware placement effects rather than inherent advantages of CVMs, we replicated the experiment on our local testbed under controlled conditions. Under identical configurations, we did not observe any performance benefit for CVMs, suggesting that the improvement in the cloud may be influenced by hardware-level factors such as NUMA locality or shared cache contention.

To examine NUMA effects, we pinned CPU and memory to different NUMA nodes on our local server (configurations 5 and 6 in Table~\ref{table: System configurations}). This configuration resulted in an average slowdown of 2.4\%, indicating the impact of cross-node memory access. We then evaluated the effect of Last-Level Cache (LLC) contention by colocating two VMs on the same socket- one executed matrix multiplication and the other ran a cache-intensive workload. Under this setup, matrix multiplication performance degraded by approximately 20\%, confirming the sensitivity of the workload to shared cache interference.

These observations suggest that variations in NUMA placement and LLC contention in multitenant cloud environments can influence performance. In practice, the lower adoption of CVMs may reduce resource contention, allowing certain workloads to experience better cache availability or locality.

\begin{figure*} [!t]
  \centering
  \includegraphics[width=1\textwidth]{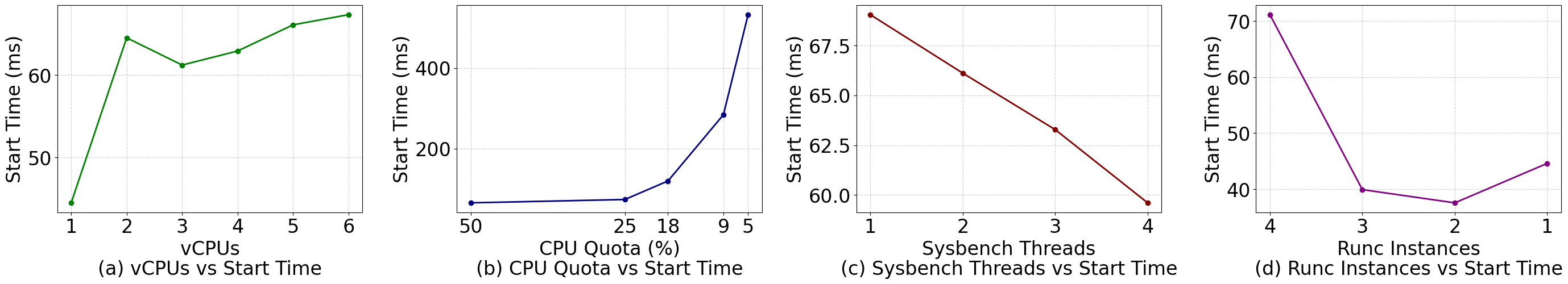}
  \caption{Runtime of runc with (a) number of vCPUs, (b) different CPU quota, (c) number of Sysbench threads, and (d) other Runc instances.}
  \label{fig:runcOnDifferentSettings}
\end{figure*}

Taken together, the factors discussed in this section primarily affect steady-state execution latency after a container has already been initialized. However, serverless responsiveness is also influenced by startup latency when new containers are instantiated. We therefore shift our focus from steady-state execution overhead to cold-start behavior in the serverless function lifecycle.

\subsection{Characterizing Cold Start Bottlenecks}
\label{subsec:cold_start}

While Section~\ref{subsec:perf_overhead} analyzes warm-start execution behavior, serverless performance is also heavily influenced by container initialization latency. In particular, cold-start overhead can dominate total function execution time, especially for short-lived workloads. To quantify this effect, we evaluated several SeBS benchmarks \cite{copik_sebs_2020} and observed that container startup accounts for around 65\% of total execution time. These findings motivate a deeper analysis of the sources of cold-start behavior.

To compare cold-start behavior, we measured container startup time in both regular VMs and CVMs. The results indicate that CVMs do not inherently introduce additional cold-start overhead relative to regular VMs. Although CVMs do not significantly increase cold-start latency, cold-start itself dominates execution time, making its optimization critical for confidential serverless deployments. We therefore investigate the factors that dominate container initialization latency across different execution configurations. We systematically analyze runc \cite{runc}, the low-level container runtime responsible for creating and launching containers, under varying CPU configurations, contention levels, and isolation mechanisms to identify the dominant contributors to cold-start latency.

\subsubsection{Single-core vs. Multi-core VMs}First, we compare running runc on a single core and multiple-core regular VMs with the hyperthreading turned off. The single vCPU execution took 44 ms while the multi-vCPU VM took over 60 ms. From Figure~\ref{fig:runcOnDifferentSettings}(a), we can see that running on a 1-vCPU VM is faster than running on a multi-vCPU VM. The increased latency in multi-vCPU VMs is attributable to additional synchronization and coordination overhead introduced by parallel execution across cores. Running on 1 core would mitigate this synchronization issue and hence make the runtime of runc faster.

\subsubsection{Impact of CPU Quota Throttling and CPU Load}Second, we see how runc performs in a condition where the cores are stressed. We ran several instances of runc on a VM by restricting the time the VM can use the CPU by implementing a quota for the VM's cgroup. Restricting the VM to use 50\% of the period did not make any impact. However, decreasing this number further increased the runtime of runc, as shown in Figure~\ref{fig:runcOnDifferentSettings}(b). 

Interestingly, when running runc alongside a CPU-intensive sysbench workload, container startup time decreased to 59 ms—approximately 6 ms faster than the baseline. This behavior arises because runc exhibits relatively low CPU utilization; under contention, the scheduler prioritizes short-lived tasks, and sustained CPU activity keeps cores active, reducing wake-up latency. As shown in Figure~\ref{fig:runcOnDifferentSettings}(c), increasing the number of sysbench threads further reduces runc startup time in multi-vCPU settings. However, in single vCPU VMs, running runc alongside a single-threaded CPU-intensive sysbench workload made runc slower. In this case, running runc alone took 44 ms, while running it alongside sysbench took 61 ms, making it slower to run with CPU-intensive workloads. Having fewer vCPUs yields higher CPU utilization of the runc container and may reduce the priority the container receives over the sysbench-CPU workload.

\subsubsection{Launching Multiple Containers in Parallel}Third, we ran 4 instances of runc on 4 different VMs. All the VMs had 4 vCPUs that were pinned to the same set of 4 cores as mentioned above. The time taken by runc on these four VMs was 54.8 ms, 55.9 ms, 56.5 ms, and 52.4 ms, which is almost 10 ms (15\%) faster than running runc solely on a 4-vCPU VM. These results indicate that the runc startup is not CPU-saturated; instead, moderate parallelism can improve performance by reducing idle periods and keeping cores active.  

From Figure~\ref{fig:runcOnDifferentSettings}(d), we can observe that running three instances of runc simultaneously on a one-vCPU VM kept the runtime of runc lower. This is because, when running runc, most of the time the CPU remains idle, and running two or three instances of runc together, in this configuration, would not impact each other, but rather keep the CPU warm, which results in better performance. However, running 4 instances throttled the CPU and hence increased the runtime.

\subsubsection{Cgroup Write Contention}
\begin{figure*}[!t]
    \centering
    \includegraphics[width=0.85\textwidth]{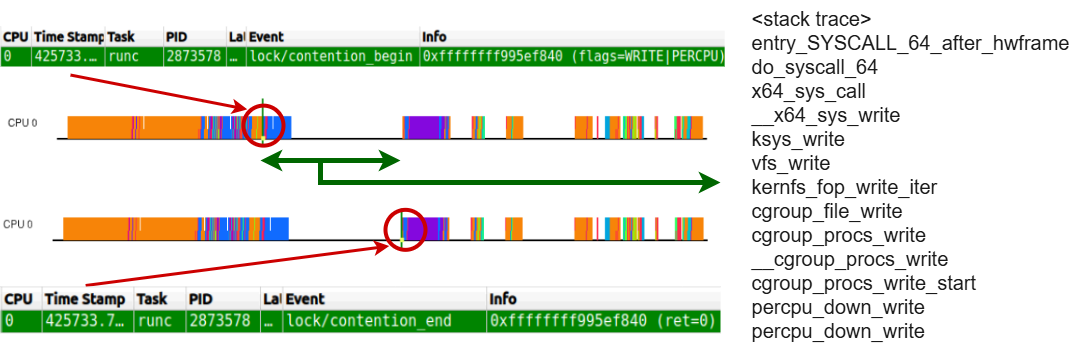}
    \caption{Contention in runc and its stack trace.}
    \label{fig:RuncContention}
\end{figure*}

Experiments above show that runc performs better in a single vCPU rather than in a multi-core environment. Forcing runc to use fewer vCPUs or running it with another workload in a controlled condition results in a lower runc runtime due to warm cores and having to suffer less contention. This contention, associated with \texttt{WRITE|PERCPU} operations during PID updates to cgroups, can sometimes account for up to 50\% of runc startup time. During this period, the runc process makes no forward progress. KernelShark \cite{Kernelshark} traces (Figure~\ref{fig:RuncContention}) show the contention interval, with the stack trace identifying \texttt{percpu\_down\_write} and the underlying \texttt{percpu\_rwsem} semaphore as the source.

\subsubsection{Reducing Launch Overhead in Trusted Settings} \label{subsubsec:trusted-launch-overhead} As discussed in Section~\ref{subsubsection:implicationsForWarmContainerCapacity}, consolidating multiple container instances of the same function within a single CVM improves memory efficiency even without cross-VM deduplication. We next consider whether selected container-level isolation mechanisms can be relaxed to reduce launch overhead under a restricted trust setting. 

\textbf{Threat Model and Security Boundary.}
We assume a per-function CVM deployment in which multiple container instances of the same serverless function execute within a dedicated CVM and are mutually trusted. Container instances belonging to different functions execute in separate CVMs, even when those functions belong to the same tenant. Containers belonging to different tenants are likewise isolated in separate CVMs. The cloud host and hypervisor remain outside the trust boundary, whereas the CVM hardware, guest OS, and container runtime are trusted. Under this model, the hardware-enforced CVM boundary provides isolation across different functions and tenants, while container-level mechanisms provide additional defense-in-depth among instances of the same function within a CVM. Relaxing selected container-level mechanisms therefore weakens isolation only among mutually trusted instances of the same function and does not remove the CVM isolation boundary protecting other functions or tenants. These optimizations should not be applied when mutually untrusted containers share an execution environment. We do not assume that CVM isolation eliminates microarchitectural side channels through shared hardware resources; evaluating such attacks is outside the scope of this work. Under this threat model, we evaluate three container-level mechanisms that contribute to container launch overhead.

\textbf{Isolations Relaxation.}
Cgroups in Linux limit the resource usage of individual processes, preventing a container from monopolizing CPU, memory, or I/O resources. We evaluate selectively removing cgroup-related operations during container creation to reduce the associated launch overhead. Similarly, Seccomp (Secure Computing Mode) restricts the system calls available to a container, reducing the kernel attack surface. We evaluate bypassing Seccomp filtering during container launch. As the threat model above shows, this optimization trades intra-CVM defense-in-depth for lower startup overhead. Finally, we examine the overhead of network namespace creation for a Docker container, which includes setting up isolated virtual network interfaces and associated network state. We configure new containers to reuse the network namespace of an already running container, thereby avoiding this setup. This optimization removes network isolation among the participating containers and is therefore considered only under the trust model defined above.

\subsection{Optimizations for Warm and Cold Start Performance}
\label{sec:ImprovingColdStart}

Based on the observations from Sections \ref{subsec:perf_overhead} and \ref{subsec:cold_start}, we summarize practical optimization strategies for both warm-start and cold-start scenarios. Some factors, such as NUMA locality and LLC contention, are largely outside user control. However, runtime configuration choices and container initialization policies offer opportunities to mitigate overhead.

In warm-start scenarios, serverless functions that rely on multithreading exhibit a trade-off between parallelism benefits and thread synchronization/VMEXITs overhead. While parallelism initially improves performance, for some serverless functions, it leads to increased synchronization and VMEXIT amplification, particularly in CVMs.

For cold starts, the observations in Section~\ref{subsec:cold_start} suggest three classes of optimization depending on their deployment requirements.

\textit{Configuration-level optimizations.}
Several optimizations can be applied through resource and runtime configuration without modifying the kernel or relaxing isolation. Our measurements show that \texttt{runc} initialization often underutilizes the available CPU resources, while excessive vCPU allocation can increase synchronization overhead. Accordingly,
selecting an appropriate number of vCPUs and controlling initialization parallelism can reduce container-launch latency.

\textit{Kernel-level optimization.}
We identify prolonged contention associated with \texttt{WRITE|PERCPU} operations on the \texttt{percpu\_rwsem} semaphore during PID writes to the cgroup hierarchy. This contention can be mitigated by enabling the \texttt{CONFIG\_CGROUP\_FAVOR\_DYNMODS} kernel configuration, which prioritizes the relevant write operations. Unlike the configuration-level optimizations above, this optimization requires a kernel configuration change and recompilation.

\textit{Trust-dependent isolation relaxation.}
Under the restricted trust model defined in Section~\ref{subsubsec:trusted-launch-overhead}, additional reductions in startup latency are possible by selectively relaxing container-level isolation mechanisms. Skipping cgroup assignment during \texttt{runc} startup reduces launch time by 27\%, bypassing Seccomp filtering
reduces startup latency by approximately 30~ms, and reusing the network namespace of an existing container reduces startup time by approximately 50~ms. These techniques trade intra-CVM isolation and defense-in-depth for lower startup latency and therefore are not general-purpose recommendations; they apply only under the trust assumptions defined in Section~\ref{subsubsec:trusted-launch-overhead}.
\vspace{-0.1 in}
\section{RELATED WORKS}

Confidential computing has emerged as a promising approach to address the risks posed by untrusted cloud infrastructures, especially in multi-tenant environments where sensitive data may be exposed to malicious insiders or compromised hypervisors. Much of the early work in this area focused on hardware-based TEEs, such as Intel SGX and ARM TrustZone, which enable isolated execution of sensitive code in secure enclaves. Prior work on SGX-based serverless computing has largely concentrated on SGX-specific limitations such as enclave memory constraints, deployment complexity, and incomplete system protection (Plug-In Enclaves \cite{pluginenclaves2021}, LLMaaS \cite{llmaas2024}, Clemmys \cite{clemmys2019}). These designs inherit the limitations of TEE-based solutions, such as the need for application refactoring, and other issues as explained in Section~\ref{subsec:SecDemand}. In contrast, our work focuses on CVM-based serverless computing, which offers promising advantages over TEE-based approaches (Section~\ref{sec:serverlessInCvm}).

Recent advances in CVMs have motivated several performance studies. Misono et al.~\cite{misono_confidential_2024} provide a comprehensive characterization of AMD SEV-SNP and Intel TDX, covering VM boot, VMEXITs, I/O, and application performance. Yan et al.~\cite{yan_performance_2023} and Qiu et al.~\cite{qiu_price_2024} similarly characterize CVM overheads using general-purpose or domain-specific workloads. These studies establish important fundamental costs of confidential execution, but do not examine how these costs manifest across the serverless execution lifecycle. In contrast, our work focuses specifically on short-lived and latency-sensitive serverless workloads. It examines three serverless-specific consequences: the impact of CVM memory inefficiency on warm-container capacity, workload-dependent warm-start execution overhead, and container runtime initialization during cold starts. We further examine both Intel TDX and AMD SEV-SNP, while conducting detailed root-cause analysis of the serverless execution overheads observed on Intel TDX.

Prior work has also directly explored confidential serverless computing, but with different objectives. Segarra et al.~\cite{coco_Segarra} present CC-Knative, a confidential deployment of the Knative serverless runtime using Confidential Containers, and primarily characterize the feasibility and substantial startup and scale-out costs introduced by CVM provisioning and attestation. Our study is complementary: rather than evaluating the construction of a confidential serverless runtime, we systematically characterize the behavior of already provisioned CVM-based serverless environments across a diverse benchmark suite, including memory efficiency, warm-start execution, and container-level cold-start bottlenecks. Sabanic et al.~\cite{sabanic2025confidentialserverlesscomputing} take a different approach by proposing a new confidential serverless architecture designed to overcome limitations of conventional CVM-based deployments, including startup latency, communication overhead, and low function density. Their evaluation therefore focuses on demonstrating the benefits of their proposed architecture. In contrast, our work is a measurement and root-cause study of conventional CVM-based serverless execution: we isolate the contribution of lost cross-VM page deduplication to memory inefficiency, connect this overhead to warm-container capacity, characterize how costly VM exits interact with serverless runtime parallelism and idle transitions, and identify container-runtime and isolation-related costs during container initialization. These analyses expose serverless-specific performance mechanisms and trade-offs that are complementary to the architectural solutions proposed in prior work.

Beyond confidential serverless systems, a broader body of work has studied performance optimization in conventional serverless platforms, including reducing cold-start latency, improving scheduling policies, using snapshot-based techniques, and exploiting container reuse and caching \cite{silva_prebaking_2020, yu_rainbowcake_2024, fuerst_faascache_2021, FaasSnap,pause_container}. Relatively few studies have explored the security implications of serverless computing, such as privacy concerns and side-channel vulnerabilities~\cite{shen_gringotts_2022}. These works target conventional platforms. Performance of serverless computing on CVMs for confidential workloads remains understudied.

\section{CONCLUSIONS}

Running sensitive serverless applications on public cloud platforms introduces privacy concerns, as the underlying host may have visibility into the VM's runtime state. Confidential Virtual Machines (CVMs) address this by encrypting memory to prevent host-level introspection. However, this added security comes with resource and performance overheads.

In this paper, we present a systematic comparison of serverless function execution across CVMs and regular VMs using fifteen different benchmarks. Our results show that memory encryption in CVMs prevents host-level page deduplication mechanisms such as Kernel Same-page Merging (KSM), resulting in reduced memory consolidation efficiency and increased memory consumption as deployment scale grows. This limitation can reduce the number of warm containers that can be maintained in memory, thereby affecting the ability of serverless platforms to mitigate cold-start latency through container reuse.

Beyond memory inefficiency, our analysis identifies key runtime factors that contribute to performance degradation in confidential serverless environments. In particular, workloads with high synchronization and idle–active transitions generate frequent VMEXITs, and this effect is amplified under CVMs due to the higher cost of VMEXIT handling. Serverless resource provisioning further exacerbates this behavior, as higher-memory configurations expose more vCPUs, encouraging increased parallelism and synchronization. In addition, reduced hypervisor visibility into guest execution limits the effectiveness of mechanisms such as Pause-Loop Exiting (PLE), making CVMs more sensitive to contention and scheduling inefficiencies.

Finally, we analyze cold-start behavior and show that container initialization can dominate overall invocation latency. Our experiments demonstrate that targeted optimizations—such as selectively relaxing unnecessary isolation mechanisms in mutually trusted same-function instances—can significantly reduce container launch time. Overall, our findings highlight the complex trade-offs between security, memory efficiency, and runtime performance in confidential serverless computing and provide practical insights for designing efficient and secure serverless platforms on emerging confidential computing infrastructures.


\bibliographystyle{IEEEtran}
\bibliography{references}

\vfill

\end{document}